\documentclass[12pt]{article}
\usepackage{epsfig}
\usepackage{color}

\newcommand{\gtap}{{\raise.3ex\hbox{$>$\kern-.75em\lower1ex\hbox{$\sim$}}}}
\newcommand{\ltap}{{\raise.3ex\hbox{$<$\kern-.75em\lower1ex\hbox{$\sim$}}}}

\begin{document}
\begin{titlepage}

\hspace*{\fill}\parbox[t]{5cm}
{hep-ph/0302168 \\
ILL-(TH)-03-02 \\
NSF-KITP-03-15 \\ \\
\today} \vskip2cm
\begin{center}
{\Large \bf Triplicated Trinification} \\
\medskip
\bigskip\bigskip\bigskip\bigskip
{\large{\bf S.~Willenbrock}} \\
\bigskip\bigskip\medskip
Department of Physics, University of Illinois at Urbana-Champaign \\
1110 West Green Street, Urbana, IL\ \ 61801 \\ \bigskip
Kavli Institute for Theoretical Physics \\University of California,
Santa Barbara, CA\ \ 93106 \\
\bigskip
\end{center}

\bigskip\bigskip\bigskip

\begin{abstract}
Gauge-coupling unification is just as successful in the standard model with
six Higgs doublets as it is in the minimal supersymmetric standard model.
However, the gauge couplings unify at $10^{14}$ GeV, which yields rapid proton
decay in the $SU(5)$ model.  I propose that the grand-unified gauge group is
instead $SU(3)_c\times SU(3)_L\times SU(3)_R$, in which baryon number is
conserved by the gauge interactions.
\end{abstract}

\end{titlepage}

Nature appears to come in triplicate.  The elementary fermions of nature come
in three identical generations of quarks and leptons, distinguished only by
their couplings to the Higgs field. We have no understanding of why nature
chooses to triplicate itself.

In contrast, there is only a single Higgs field in the standard model, which
is responsible for breaking the electroweak symmetry and generating the masses
of all the particles.  This is the simplest model of electroweak symmetry
breaking, and it is consistent with all data.  However, it seems odd that
there should be only one Higgs field, when the fermion fields come in
triplicate.

The gauge sector of the standard model, based on the symmetry $SU(3)_c\times
SU(2)_L\times U(1)_Y$, does not come in triplicate.  It was observed long ago
that this gauge symmetry can be unified into an $SU(5)$ gauge group with a
single gauge coupling \cite{Georgi:sy}.  When the $SU(5)$ gauge symmetry is
broken at a high energy scale, the $SU(3)_c\times SU(2)_L\times U(1)_Y$ gauge
couplings evolve to their low-energy values \cite{Georgi:yf}.  However,
precision measurements reveal that the low-energy values of the gauge
couplings are not consistent with $SU(5)$ grand unification.

As is well known, the minimal supersymmetric standard model nudges the
relative evolution of the gauge couplings just enough to bring them into
accord with $SU(5)$ grand unification
\cite{Ellis:1990wk,Amaldi:1991cn,Langacker:1991an}.  The reason for this is
three-fold.  First, the relative evolution of the gauge couplings is unaffected
when one adds a complete $SU(5)$ representation \cite{Georgi:yf}.  Since the
fermions are in complete $SU(5)$ representations, the addition of their
superpartners does not affect the relative evolution of the gauge couplings
\cite{Dimopoulos:1981yj}.  Second, the superpartners of the gauge bosons
(which are not in complete $SU(5)$ representations) change only the unification
scale, since they have the same gauge structure as the gauge bosons
\cite{Dimopoulos:1981yj}. Third, the minimal supersymmetric standard model
requires two Higgs doublets in order to generate masses for all the fermions.
Since the Higgs field is not in a complete $SU(5)$ representation, the
addition of a second Higgs doublet, as well as the superpartners of these two
Higgs doublets, modifies the relative evolution of the gauge couplings. Thus
it is the extension of the Higgs sector that is behind the successful $SU(5)$
unification of the gauge couplings in the minimal supersymmetric standard
model \cite{Einhorn:1981sx,Marciano:1981un}.

In the renormalization-group equations responsible for the evolution of the
gauge couplings, a (chiral) fermion field counts twice as much as a (complex)
scalar field with the same gauge quantum numbers, at least at leading order.
Thus the successful $SU(5)$ unification of the gauge couplings in the minimal
supersymmetric standard model, with its two Higgs doublets and their fermionic
superpartners, can be mimicked by the standard model with six Higgs doublets.
Since six is a multiple of three, this implies a triplication of the Higgs
sector, in keeping with the triplication of the fermion fields.

\begin{figure}
\begin{center}
\input{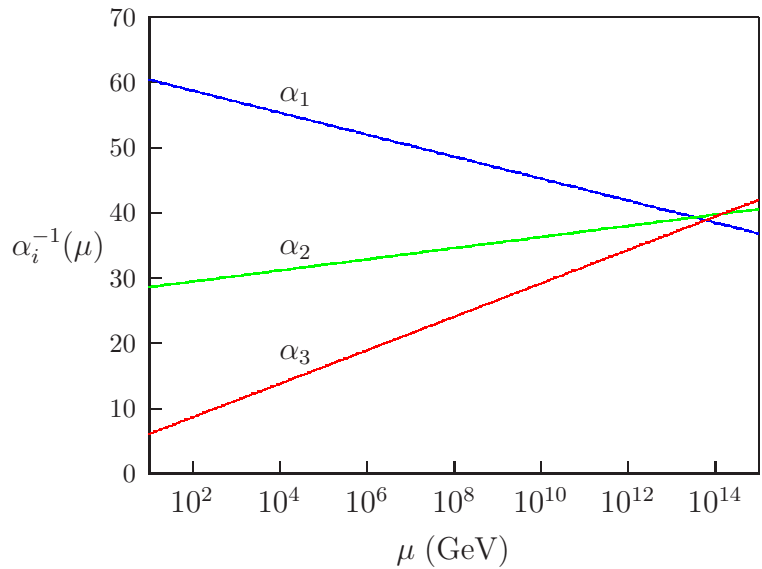}
\end{center}
\caption{Leading-order evolution of the gauge couplings from their low-energy
values to the unification scale in the six-Higgs-doublet standard model.  The
couplings meet around $10^{14}$ GeV, within the accuracy of a leading-order
calculation.} \label{fig:evolution}
\end{figure}

Thus, with respect to the unification of gauge couplings, the
six-Higgs-doublet standard model is on the same footing as the minimal
supersymmetric standard model.  However, the unification of the gauge
couplings occurs at a lower scale in the six-Higgs-doublet model.  The
evolution of the gauge couplings is given at leading order by
\begin{equation}
\frac{1}{\alpha_i(\mu)}-\frac{1}{\alpha_i(\mu')}=\frac{b_i}{2\pi}
\ln\left(\frac{\mu'}{\mu}\right)\;,\label{rng}
\end{equation}
where $b_n$ are the one-loop beta-function coefficients,
\begin{equation}
b_n=-\frac{11}{3}C_2(G)+\frac{2}{3}\sum_{\rm fermions} T(R)
+\frac{1}{3}\sum_{\rm scalars} T(R)\;,\label{beta}
\end{equation}
where $C_2(SU(N))=N$, $T(R) = 1/2$ for fermions and scalars in the fundamental
or antifundamental representation of $SU(N)$, and $C_2(U(1))=0$, $T(R)={3\over
5}Y^2$ for fermions and scalars of hypercharge $Y$ (in the convention
$Q=T_{3L}+Y$). Equation~(\ref{beta}) shows that (chiral) fermions count twice
as much as (complex) scalars in the evolution of the couplings, as noted above.
The unification scale, $M_U$, may be obtained from the condition
$\alpha_3(M_U)=\alpha_2(M_U)$, which gives
\begin{equation}
\frac{1}{\alpha_3(M_Z)}-\frac{1}{\alpha_2(M_Z)}=\frac{b_3-b_2}{2\pi}
\ln\left(\frac{M_U}{M_Z}\right)\;.\label{Mu}
\end{equation}
For $N_H$ Higgs doublets, $b_3-b_2=-11/3-N_H/6$.  Using as inputs
$\alpha_3(M_Z) = 0.117$ and $\alpha_2(M_Z) = (\sqrt 2/\pi)G_FM_W^2 = 0.034$,
Eq.~(\ref{Mu}) yields $M_U \approx 10^{14}$ GeV for $N_H=6$.  This is much less
than the unification scale of supersymmetric $SU(5)$, $M_U\approx 2 \times
10^{16}$ GeV \cite{Ellis:1990wk,Amaldi:1991cn,Langacker:1991an}. The unified
gauge coupling, obtained from Eq.~(\ref{rng}), is
$\alpha_U(M_U)=\alpha_3(M_U)= 0.025$, which is close to the value in the
original non-supersymmetric $SU(5)$ model \cite{Langacker:1980js}.

I show in Fig.~\ref{fig:evolution} the evolution of the gauge couplings from
their low-energy values up to the unification scale, using Eq.~(\ref{rng}).
The input value of the hypercharge coupling is
\begin{equation}
\alpha_1(M_Z) = (5/3)\alpha_2(M_Z)\tan^2\theta_W=0.017\;,
\label{hypercharge}
\end{equation}
where the factor $5/3$ is determined by the embedding of $U(1)_Y$ in $SU(5)$.
The couplings meet around $10^{14}$ GeV, within the accuracy of a
leading-order calculation.

A unification scale as low as $10^{14}$ GeV is disastrous --- it yields rapid
proton decay via the exchange of $SU(5)$ gauge bosons. Therefore, the
six-Higgs-doublet model cannot be embedded in a conventional $SU(5)$ theory.
One must modify the theory in some way to adequately suppress proton
decay.\footnote{The minimal supersymmetric $SU(5)$ model is also tightly
constrained by proton decay, due to one-loop processes involving superpartners
\cite{Goto:1998qg,Murayama:2001ur,Bajc:2002bv,Bajc:2002pg}.}

Rather than following this tack, I consider the possibility that the unified
theory is something other than $SU(5)$.  In general, a unified theory based on
a simple group leads to proton decay \cite{Gell-Mann:1976pg}.  However, a
product group, supplemented with a discrete symmetry to enforce the equality
of the gauge couplings, need not contain gauge bosons that mediate proton
decay. The simplest theory of this type that has the same condition on the
gauge couplings at the unification scale as $SU(5)$ is $SU(3)_c\times
SU(3)_L\times SU(3)_R$, supplemented with a discrete cyclic symmetry $Z_3$
that acts on the three groups.  The condition on the gauge couplings at the
unification scale is the same in these two theories because they are both
subgroups of $E_6$.  The weak $SU(2)_L$ group is a subgroup of $SU(3)_L$, and
$U(1)_Y$ is a linear combination of $U(1)$ subgroups contained in $SU(3)_L$ and
$SU(3)_R$. This is the so-called ``trinified'' model \cite{deRujula,Babu:gi}.
Remarkably, the triplication of the Higgs sector leads us to the triplication
of the gauge sector.

The fermions of one generation are contained in the 27-dimensional
representation of $SU(3)_c\times SU(3)_L\times SU(3)_R$,
\begin{equation}
27=(3,\bar 3,1)+(\bar 3,1,3)+(1,3,\bar 3)\;.
\end{equation}
Quarks are contained in the $(3,\bar 3,1)$ representation, antiquarks in the
$(\bar 3,1,3)$ representation, and leptons and antileptons in the $(1,3,\bar
3)$ representation.  Thus baryon number is automatically conserved by the gauge
interactions.  The 27-dimensional representation decomposes under
$SU(3)_c\times SU(2)_L\times U(1)_Y$ as
\begin{equation}
27=2(1,1,0)+(1,2,{1\over 2})+(3,1,-{1\over 3})+2[(1,2,-{1\over 2})+(\bar
3,1,{1\over 3})]+(1,1,1)+(\bar 3,1,-{2\over
3})+(3,2,{1\over6})\;.\label{decomp}
\end{equation}
This contains the familiar quarks and leptons, as well as additional fermions
in the representations $2(1,1,0),(1,2,{1\over 2}),(1,2,-{1\over
2}),(3,1,-{1\over 3}),(\bar 3,1,{1\over 3})$. These additional fermions are an
unattractive feature, but since they are in a vectorlike representation of
$SU(3)_c\times SU(2)_L\times U(1)_Y$ they naturally acquire masses of order
$M_U$ when the unification-scale symmetry breaking occurs
\cite{Georgi:1979md}.  Therefore they do not influence the evolution of the
gauge couplings, unless their masses are less than $M_U$. The two gauge-singlet
fermions (dubbed ``neutrettos'' \cite{deRujula}, although ``sterile
neutrinos'' is also appropriate) do not affect the evolution of the gauge
couplings regardless of their masses.  Indeed, in the simplest models
(discussed below), the two neutrettos acquire mass at one loop and are
therefore lighter than $M_U$ \cite{deRujula,Babu:gi}.  The other additional
fermions do not affect the relative evolution of the gauge couplings if their
masses are less than $M_U$ and nearly degenerate, because they fill out a
complete $5+\bar 5$ representation of $SU(5)$.

The details of the model depend upon on how the $SU(3)_c\times SU(3)_L\times
SU(3)_R$ symmetry is broken, first to $SU(3)_c\times SU(2)_L\times U(1)_Y$ and
then to $SU(3)_c\times U(1)_{EM}$, and on how this symmetry breaking is
communicated to the fermions.  Let us adopt the standard four-dimensional
weakly-coupled field-theory framework in which the symmetry breaking is
accomplished via the vacuum-expectation values of Higgs fields.  In this
framework, the symmetry breaking is communicated to the fermions via Yukawa
couplings.

The simplest model that yields the desired pattern of symmetry breaking
consists of two Higgs fields in the 27-dimensional representation
\cite{deRujula,Babu:gi}. However, in keeping with the triplication paradigm, it
is natural to expect three or six such Higgs fields. Each 27-dimensional
representation contains one $SU(3)_c\times SU(2)_L\times U(1)_Y$
representation with the quantum numbers of the standard-model Higgs field and
two with the quantum numbers of the conjugate Higgs field [see
Eq.~(\ref{decomp})]. As far as the evolution equations are concerned, these
representations contribute equally. Whether there are two, three, or six
27-dimensional-representation Higgs fields, one must arrange for six of the
Higgs doublets to have masses of order the weak scale and all the remaining
$SU(3)_c\times SU(2)_L\times U(1)_Y$ Higgs representations to have masses of
order the unification scale. How this occurs depends upon the details of the
symmetry-breaking sector.  This evokes the usual gauge hierarchy problem: Why
are some Higgs fields much lighter than others?  Furthermore, since the model
is non-supersymmetric, there is no mechanism to cancel quadratic divergences
in the radiative corrections to the masses of the six weak-scale Higgs
doublets.

Fermions acquire mass by coupling to these 27-dimensional Higgs fields. In
general, these Yukawa couplings mediate proton decay via Higgs-boson
exchange.  However, it is possible to impose baryon number as a global
symmetry.  There are two types of Yukawa couplings allowed
\cite{deRujula,Babu:gi},
\begin{equation}
\psi(3,\bar 3,1)\psi(\bar 3,1,3)\phi(1,3,\bar 3)+\psi(\bar 3,1,3)\psi(1,3,\bar
3)\phi(3,\bar 3,1)+\psi(1,3,\bar 3)\psi(3,\bar 3,1)\phi(\bar 3,1,3)
\end{equation}
and
\begin{equation}
\psi(1,3,\bar 3)\psi(1,3,\bar 3)\phi'(1,3,\bar 3)+\psi(3,\bar 3,1)\psi(3,\bar
3,1)\phi'(3,\bar 3,1)+\psi(\bar 3,1,3)\psi(\bar 3,1,3)\phi'(\bar 3,1,3)\;.
\end{equation}
We assign to the fermion fields the canonical baryon numbers of $1/3$ to
$\psi(3,\bar 3,1)$, $-1/3$ to $\psi(\bar 3,1,3)$, and zero to $\psi(1,3,\bar
3)$.  In the first Yukawa coupling, one may assign baryon number zero to
$\phi(1,3,\bar 3)$, $1/3$ to $\phi(3,\bar 3,1)$, and $-1/3$ to $\phi(\bar
3,1,3)$.  In the second Yukawa coupling, the assignments are zero to
$\phi'(1,3,\bar 3)$, $-2/3$ to $\phi'(3,\bar 3,1)$, and $2/3$ to $\phi'(\bar
3,1,3)$.  If all 27-dimensional Higgs fields participate in only one Yukawa
coupling or the other, and if the Higgs potential is constrained to conserve
baryon number, then proton decay is absolutely forbidden
\cite{deRujula,Babu:gi}.

In the minimal model, with just two 27-dimensional Higgs fields, it is not
possible to impose baryon number and to generate a realistic fermion mass
spectrum and Cabibbo-Kobayashi-Maskawa (CKM) mixing matrix.  To conserve baryon
number, one of these Higgs fields must participate in the first Yukawa
coupling, the other in the second Yukawa coupling.  With just one Higgs field
participating in the first Yukawa coupling, the up and down quark mass
matrices are proportional, yielding a unit CKM matrix and
$m_u/m_d=m_c/m_s=m_t/m_b$ \cite{deRujula,Babu:gi}.  Thus baryon number is
necessarily violated in the minimal model. However, with three or more
27-dimensional Higgs representations, the imposition of baryon number symmetry
is possible \cite{deRujula,Babu:gi}.

Although it is possible for the Yukawa couplings to conserve baryon number,
this is a stronger requirement than is necessary.   Since the Yukawa couplings
of the first two generations (relevant for proton decay) are much smaller than
gauge couplings, one can tolerate some Higgs-mediated baryon-number violation
while respecting the lower bound on the proton lifetime.

The challenge is to find a model that fits nature, and to extract its
predictions for physics beyond the standard model.  In building such a model,
there are pitfalls beyond proton decay to avoid.  Models with multiple Higgs
doublets generically have tree-level Higgs-mediated flavor-changing neutral
currents \cite{Glashow:1976nt}, which are severely constrained
experimentally.   However, even if tree-level flavor-changing neutral currents
are present, they may be sufficiently suppressed by small Yukawa couplings to
avoid conflicting with experiment \cite{Sher:1991km,Antaramian:1992ya},
analogous to the suppression of Higgs-mediated proton decay mentioned above.

With the triplication of both the fermion and the Higgs sectors, it is
tempting to look for family symmetries relating the fermion fields as well as
the Higgs fields.  A model with six Higgs doublets has recently been explored
in Ref.~\cite{Adler:1998as,Adler:1999gv}. The approximate unification of gauge
couplings was also noted in that work. The difficulties with Higgs-mediated
flavor-changing neutral currents discussed above are exemplified in that study.

The unification scale of $10^{14}$ GeV is several orders of magnitude below
the reduced Planck scale, $(8\pi G_N)^{-1/2}\approx 2\times 10^{18}$ GeV.  In
weakly-coupled string theory, the string scale cannot be much less than the
reduced Planck scale.  Thus in this scenario, the model is an $SU(3)_c\times
SU(3)_L\times SU(3)_R$ field theory between the grand-unified and string
scales.  This gauge group is a maximal subgroup of $E_6$, so it naturally
arises in string theory \cite{Witten:1985xc,Greene:1986ar}.  Furthermore, the
fermion and Higgs representations arise at Kac-Moody level one, which
corresponds to the simplest string models.  The $SU(3)_c\times SU(3)_L\times
SU(3)_R$ model may be embedded into a larger gauge group, such as $E_6$
\cite{Gursey:1975ki}, but this must be broken well above $10^{14}$ GeV in
order to avoid rapid proton decay via gauge-boson exchange. A Higgs field in
the 650-dimensional representation of $E_6$ could provide the desired breaking
to $SU(3)_c\times SU(3)_L\times SU(3)_R$.

In strongly-coupled string theory, the string scale can be much less than the
Planck scale \cite{Witten:1996mz}.  Thus it is possible that the string scale
and the unification scale are both $10^{14}$ GeV.  This opens up additional
possibilities for model building.  If all three $SU(3)$ gauge symmetries are
realized at Kac-Moody level one (or, more generally, at the same Kac-Moody
level), then conformal symmetry equates their couplings
\cite{Ginsparg:1987ee}, and obviates the discrete cyclic $Z_3$ symmetry. This
is superior to string unification of the standard-model gauge group, where the
normalization of the $U(1)_Y$ coupling may take any rational value, not
necessarily the value $5/3$ needed for gauge-coupling unification [see
Eq.~(\ref{hypercharge})]. There are also additional possibilities for
grand-unified symmetry breaking based on compactification of extra dimensions
\cite{Witten:1985xc,Candelas:en}. Similar mechanisms also operate in
extra-dimensional field theories \cite{Kawamura:2000ev, Hall:2001pg}.

A firm prediction of triplicated trinification, independent of the details of
the model, is a weak-scale spectrum consisting of the standard model with six
Higgs doublets. The model cannot be supersymmetric, because the fermionic
superpartners of the six Higgs doublets would upset the unification of the
couplings.  Experiments at the Fermilab Tevatron and the CERN Large Hadron
Collider should see some or all of the particles contained in these six Higgs
doublets (eleven neutral scalars, five pairs of singly-charged scalars).

Should this model prove to be correct, it would mean that the supersymmetric
SU(5) model is a red herring.  Both the six-Higgs-doublet standard model and
the supersymmetric standard model yield successful gauge-coupling unification
due to the extension of the Higgs sector.  The fermion content of the standard
model fits perfectly into the $\bar 5+10$ of $SU(5)$, while the trinification
model requires additional fermions to fill out the fermion representation.
However, these additional fermions are in a vectorlike representation of the
standard-model gauge group, so it is natural for them to have masses of order
the grand-unified scale.   Thus the triplicated trinification model is a
natural extension of the standard model.

\section*{Acknowledgments}

\indent\indent  I am grateful for conversations with C.~Albright, A.~de
Gouv\^ea, J.~Polchinski, I.~Rothstein, and R.~Volkas.  This research was
supported in part by the U.~S.~Department of Energy under contract
No.~DE-FG02-91ER40677 and by the National Science Foundation under Grant
No.~PHY99-07949.


\end{document}